\documentclass{article}
\usepackage{spconf,amsmath,graphicx}
\usepackage{hyperref}
\usepackage{caption}
\usepackage{algorithm,algorithmicx,algpseudocode}

\usepackage{amsmath,amsfonts,amssymb,pifont}
\usepackage{adjustbox}
\usepackage{comment}

\title{Learning Speech Representation From Contrastive Token-Acoustic Pretraining}

\name{Chunyu Qiang$^{1,2,3}$, Hao Li$^{2}$, Yixin Tian$^{2}$, Ruibo Fu$^{4}$, Tao Wang$^{4}$, Longbiao Wang$^{1,3,*}$, Jianwu Dang$^{3}$}

\address{$^1$School of New Media and Communication, Tianjin University, Tianjin, China \\
$^2$Kuaishou Technology Co., Ltd, Beijing, China \\
$^3$Tianjin Key Laboratory of Cognitive Computing and Application, \\
College of Intelligence and Computing, Tianjin University, Tianjin, China \\
$^4$Institute of Automation, Chinese Academy of Sciences, Beijing, China }

\begin{document}
\ninept
\maketitle

\begin{abstract}
For fine-grained generation and recognition tasks such as minimally-supervised text-to-speech (TTS), voice conversion (VC), and automatic speech recognition (ASR), the intermediate representations extracted from speech should serve as a "bridge" between text and acoustic information, containing information from both modalities. The semantic content is emphasized, while the paralinguistic information such as speaker identity and acoustic details should be de-emphasized. However, existing methods for extracting fine-grained intermediate representations from speech suffer from issues of excessive redundancy and dimension explosion. Contrastive learning is a good method for modeling intermediate representations from two modalities. However, existing contrastive learning methods in the audio field focus on extracting global descriptive information for downstream audio classification tasks, making them unsuitable for TTS, VC, and ASR tasks. To address these issues, we propose a method named "Contrastive Token-Acoustic Pretraining (CTAP)", which uses two encoders to bring phoneme and speech into a joint multimodal space, learning how to connect phoneme and speech at the frame level. The CTAP model is trained on 210k speech and phoneme pairs, achieving minimally-supervised TTS, VC, and ASR. The proposed CTAP method offers a promising solution for fine-grained generation and recognition downstream tasks in speech processing. We provide a website with audio samples. \href{https://qiangchunyu.github.io/CPSP/}{$^1$}
\end{abstract}

\renewcommand{\thefootnote}{\fnsymbol{footnote}} 
\footnotetext{$*$ Corresponding author.} 
\footnotetext{Audio samples: https://qiangchunyu.github.io/CPSP/}


\begin{figure*}[t]
 \centering
 \includegraphics[width=\linewidth]{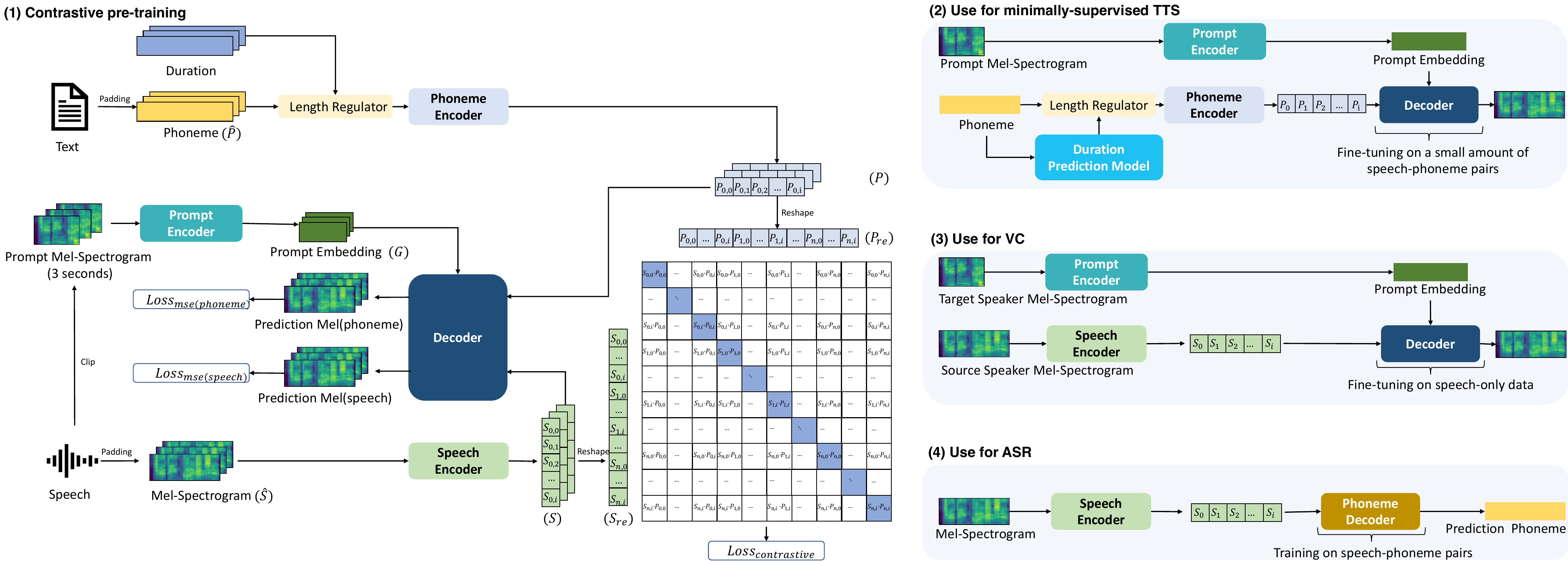}
 \vspace{20pt}
 \captionsetup{belowskip=10pt}
 \caption{The CTAP jointly trains a speech encoder, a phoneme encoder, a prompt encoder, and a decoder using contrastive learning to learn frame-level (dis)similarity between a batch of speech$(\hat{S})$ and phoneme$(\hat{P})$ pairs. During inference, the pre-trained encoders are used to extract speech embeddings$(S)$ from the input speech and phoneme embeddings$(P)$ from the input text. The CTAP can be used for downstream tasks involving frame-level generation and recognition (minimally-supervised TTS, VC, ASR, etc.).}
 \label{fig:proposed_model}
\end{figure*}

\section{Introduction}
\label{sec:intro}

Deep learning approaches have brought about a significant improvement in the field of speech representation, as evidenced by the remarkable performance achieved\cite{mohamed2022self}. Speech processing encompasses various tasks. For fine-grained generation and recognition tasks like TTS, VC, and ASR, the intermediate representation extracted from speech should serve as a "bridge" between text and acoustic information. It should emphasize linguistic content while de-emphasizing paralinguistic information such as speaker identity and acoustic details. Consequently, developing a suitable representation learning model for TTS, VC, and ASR poses a challenge. 

Self-supervised representation learning methods, such as \\ Wav2Vec2.0\cite{baevski2020wav2vec}, Wav2Vec-C\cite{sadhu2021wav2vec}, VQ-Wav2Vec\cite{baevski2019vq}, HuBERT\cite{hsu2021hubert}, and W2V-BERT\cite{chung2021w2v}, offer the prospect of a universal model that can benefit a wide range of tasks and domains. While these methods can be applied to tasks such as ASR, they encounter issues of redundancy and dimension explosion when it comes to VC task and minimally-supervised TTS task like SPEAR-TTS\cite{kharitonov2023speak} and Diff-LM-Speech\cite{qiang2023minimally}. 

In supervised representation learning methods, the PPGs\cite{sun2016phonetic} are computed from the ASR acoustic model. Although widely used in VC tasks, it is essentially textual information and cannot be used in minimally-supervised TTS task.

Contrastive models address these challenges by learning representations through differentiating a target sample (positive) from distractor samples (negatives) based on an anchor representation. The objective is to maximize the similarity between the anchor and positive samples while minimizing the similarity between the anchor and negative samples. This approach has been widely used in computer vision, exemplified by Open AI's CLIP\cite{radford2021learning}, Florence\cite{yuan2021florence}, and ALIGN\cite{jia2021scaling}. In the audio field, CLIP-based models like Wav2CLIP\cite{wu2022wav2clip}, AudioCLIP\cite{guzhov2022audioclip}, and CLAP\cite{elizalde2023clap} have been developed. However, these methods focus on extracting global descriptive information from audio for downstream audio classification tasks, making them inadequate for fine-grained generation and recognition tasks like TTS, VC, and ASR.

To tackle these challenges, we propose a method called "Contrastive Token-Acoustic Pretraining (CTAP)". The CTAP utilizes two encoders to bring phoneme and speech into a joint multimodal space, facilitating the connection between phoneme and speech at the frame level. Our contributions in this paper are as follows:

\begin{itemize}
\item Introducing the CTAP model trained on 210k speech and phoneme pairs.

\item Achieving minimally-supervised TTS by fine-tuning the decoder using a small amount of labeled data.

\item Simultaneously achieving VC using non-parallel speech-only data.

\item Implementing ASR, which can be generalized to multiple fine-grained generation and recognition downstream tasks.

\end{itemize}


\section{Method}

\subsection{Contrastive Token-Acoustic Pretraining}
\subsubsection{Architecture}
CTAP is illustrated in Fig. \ref{fig:proposed_model}. CTAP mainly consists of four components, including a speech encoder, a phoneme encoder, a prompt encoder, and a decoder. The length regulator is used to solve the problem of length mismatch between phoneme sequences and speech sequences. A duration diffusion model\cite{qiang2023minimally} is used. The input is speech and phoneme pairs passed to a speech encoder and a phoneme encoder. Both representations are connected in joint multimodal space. The space is learned with the frame-level (dis)similarity of speech and phoneme pairs in a batch using contrastive learning. This method is inspired by the CLIP model\cite{radford2021learning} and the CLAP model \cite{elizalde2023clap}.

\subsubsection{Training Details}
$\hat{S}$ denote the input batch of speech data, $\hat{S} \in \mathbb{R}^{B \times T_s \times D_s}$ where $B$ is the batch size, $T_s$ is the number of time bins, and $D_s$ is the number of spectral components (e.g. Mel bins). $\hat{P}$ denotes the input batch of phoneme data, $\hat{P} \in \mathbb{R}^{B \times T_p \times D_p}$ where $T_p$ is the number of phoneme codes, and $D_p$ is the number of phoneme coding dimensions. During training, the ground-truth duration is used to expand the phoneme sequence's length to $T_s$. During inference, the corresponding predicted duration is used. The phoneme sequence after the length regulator is obtained: $\hat{P} \in \mathbb{R}^{B \times T_s \times D_p}$. From the pairs, the speech and phoneme are passed through a speech encoder and a phoneme encoder respectively:

\begin{equation}
\begin{aligned}
 {S} = Encoder_s(\hat{S});
 {P} = Encoder_p(\hat{P})
\end{aligned}
\end{equation}

where ${S} \in \mathbb{R}^{B \times T_s \times d}$ are the speech representations, and ${P} \in \mathbb{R}^{B \times T_s \times d}$ are the phoneme representations. We bring speech and phoneme representations, ${S}$ and ${P}$, into a joint multimodal space of dimension $d$.

The prompt speech is randomly clipped with a window of length 3 seconds at each step as the input to the prompt encoder. $G$ denotes the prompt embedding. $G \in \mathbb{R}^{B \times D_g}$ where $D_g$ is the number of dimensions. The prompt encoder is a VAE-based model \cite{qiang2023improving} that extracts paralinguistic information to address the one-to-many problem in text-to-speech synthesis, such as timbre, style, and prosody, from the prompt speech. To address the KL collapse problem, a margin $\Delta$ is introduced to limit the minimum value of the KL loss as shown: 

\begin{equation}
\begin{aligned}
 \mathcal{L}_{kl} = max(0, D_{KL}[\mathcal{N}({\mu},{\sigma}^2)||\mathcal{N}(0, I)]-\Delta
\end{aligned}
\end{equation}

Both$S$ and$P$ are input to the same $Decoder$ to predict mel-spectrograms. 

\begin{equation}
\begin{aligned}
 Mel_{s} = Decoder(S, G);
 Mel_{p} = Decoder(P, G)
\end{aligned}
\end{equation}

The predicted mel-spectrograms are compared with the ground-truth mel-spectrograms using mean square error (MSE) loss.

\begin{equation}
     \mathcal{L}_{mse}= 0.5 * (\ell_{mse}(Mel_{s}) + \ell_{mse}(Mel_{p}))
\end{equation}

To extract frame-level representation,$S$ and$P$ within a batch are reshaped into 2D matrixs$S_{re}$ and$P_{re}$, where$S_{re}$ and$P_{re} \in \mathbb{R}^{(B * T_s) \times d}$. This approach is beneficial for contrastive learning, as it increases the number of sample pairs per step. 
Now that the$S_{re}$ and$P_{re}$ are comparable, we can measure similarity:

\begin{equation}
    C = \tau*({S}_{re} \cdot {P}_{re}^\top)
\end{equation}

where $\tau$ is a temperature parameter to scale the range of logits. The similarity matrix $C \in \mathbb{R}^{(B * T_s) \times (B * T_s)}$ has $(B * T_s)$ correct pairs in the diagonal and $(B * T_s)^2-(B * T_s)$ incorrect pairs in the off-diagonal. As the extracted intermediate representation contains contextual information, only the current frame corresponds to a positive sample.

\begin{equation}
     \mathcal{L}_{contrastive}= 0.5 * (\ell_{speech}(C) + \ell_{phoneme}(C))
\end{equation}

where $\ell_{k} = \frac{1}{B * T_s}\sum_{i=0}^{B * T_s} \log diag (softmax(C))$ along speech and phoneme axis respectively. We used this symmetric cross-entropy loss ($\mathcal{L}_{contrastive}$) over the similarity matrix to jointly train the speech encoder and the phoneme encoder.
The total loss of the model is: 

\begin{equation}
\begin{aligned}
 \mathcal{L} = \mathcal{L}_{contrastive} + \mathcal{L}_{mse} + \mathcal{L}_{kl}
 \label{L}
\end{aligned}
\end{equation}

\subsection{Application}

\subsubsection{Minimally-supervised TTS}
\label{sec:tts}
Minimally-supervised TTS refers to splitting the TTS task into two tasks (text-to-semantic and semantic-to-speech) by combining speech intermediate representations. For the minimally-supervised TTS task, as shown in Fig. \ref{fig:proposed_model}, the weights of the pre-trained phoneme encoder and speech encoder are frozen. The target speaker's speech-only data is used to extract speech embeddings through the speech encoder, which is used to fine-tune the decoder and prompt encoder. The decoder is then fine-tuned again with a small amount of speech-phoneme pairs. Pre-trained phoneme encoder is used to extract phoneme embeddings. During inference, phoneme is input into the phoneme encoder to obtain phoneme embeddings, and prompt speech is input into the prompt encoder to obtain prompt embeddings. The phoneme embeddings and prompt embeddings are then fed into the decoder to generate predicted mel-spectrograms.

\subsubsection{VC}
\label{sec:vc}
Similarly, for the VC task, parallel data is not required. Instead, the decoder and prompt encoder are fine-tuned using the target speaker's speech-only data. During inference, speech is input into the speech encoder to obtain speech embeddings, and prompt speech is input into the prompt encoder to obtain prompt embeddings. The speech embeddings and prompt embeddings are then fed into the decoder to generate predicted mel-spectrograms.

\subsubsection{ASR}
\label{sec:asr}
In the ASR task, we trained a phoneme decoder to predict phonemes from speech embeddings. Pre-trained speech encoder is used to extract speech embeddings.

\section{Experiments}

\begin{table*}[]
\captionsetup{skip=20pt} 
\vspace{20pt}
 \caption{Results of Minimally-Supervised TTS, VC, and ASR}
 \label{tab:mos}
 \centering
\resizebox{\linewidth}{!}{ 
\begin{tabular}{l|ccccc|cccc|c}
\hline
\multicolumn{1}{c|}{{Model}} & \multicolumn{5}{c|}{Minimally-Supervised TTS Task}                                                                                        & \multicolumn{4}{c|}{VC Task}                                                                                     & ASR Task                                                                                  \\ \cline{2-11} 
\multicolumn{1}{c|}{}                       & {MSEP} & \multicolumn{1}{c|}{{WER}} & \multicolumn{3}{c|}{MOS}                                              & \multicolumn{1}{c|}{{WER}} & \multicolumn{3}{c|}{MOS}                                             & {\begin{tabular}[c]{@{}c@{}}ACC\end{tabular}} \\ \cline{4-6} \cline{8-10}
\multicolumn{1}{c|}{}                       &                       & \multicolumn{1}{c|}{}                     & Prosody Sim           & Speaker Sim           & Speech Quality         & \multicolumn{1}{c|}{}                     & Prosody Sim           & Speaker Sim          & Speech Quality         &                                                                                           \\ \hline
Codec\cite{Defossez2022HighFN}                                       & 99.0                  & \multicolumn{1}{c|}{6.2}                  & 3.61 ± 0.044          & 3.74 ± 0.040          & 3.68 ± 0.087          & \multicolumn{1}{c|}{4.9}                  & 3.99 ± 0.050          & 2.94 ± 0.021         & 3.56 ± 0.009          & \textbackslash{}                                                                          \\ \hline
Wav2Vec2.0\cite{baevski2020wav2vec}                                  & 102.2                 & \multicolumn{1}{c|}{4.9}                  & 3.80 ± 0.060          & 3.70 ± 0.019          & 3.90 ± 0.009          & \multicolumn{1}{c|}{4.3}                  & 4.02 ± 0.091          & 3.54 ± 0.002         & 3.57 ± 0.054          & 93.28                                                                                     \\ \hline
Hubert\cite{hsu2021hubert}                                      & 99.8                  & \multicolumn{1}{c|}{4.7}                  & 3.89 ± 0.013          & 3.71 ± 0.077          & 3.83 ± 0.002          & \multicolumn{1}{c|}{\textbf{4.2}}         & 4.00 ± 0.090          & 3.44 ± 0.039         & 3.94 ± 0.030          & \textbf{95.99}                                                                            \\ \hline
Whisper Encoder\cite{radford2023robust}                             & \textbackslash{}      & \multicolumn{1}{c|}{\textbackslash{}}     & \textbackslash{}      & \textbackslash{}      & \textbackslash{}      & \multicolumn{1}{c|}{5.0}                  & 3.98 ± 0.012          & 3.02 ± 0.058         & 3.76 ± 0.074          & 95.89                                                                                     \\ \hline
CTAP(Groud Truth)                           & 78.9                  & \multicolumn{1}{c|}{2.8}                  & 4.32 ± 0.098          & 4.53 ± 0.057          & 4.19 ± 0.017          & \multicolumn{1}{c|}{\textbackslash{}}     & \textbackslash{}      & \textbackslash{}     & \textbackslash{}      & \textbackslash{}                                                                          \\ \hline
CTAP w/o $decoder$                            & 101.2                 & \multicolumn{1}{c|}{7.2}                  & 3.88 ± 0.039          & 3.70 ± 0.079          & 3.89 ± 0.005          & \multicolumn{1}{c|}{5.1}                  & 3.99 ± 0.030          & 3.59 ± 0.009         & 3.80 ± 0.080          & \textbackslash{}                                                                          \\ \hline
CTAP + $\mathcal{L}_{mse(embed)}$                               & \textbf{98.1}         & \multicolumn{1}{c|}{4.5}                  & 3.91 ± 0.078 & 3.93 ± 0.050          & 3.93 ± 0.030          & \multicolumn{1}{c|}{4.3}                  & 4.03 ± 0.043          & 3.85 ± 0.077         & \textbf{3.95 ± 0.055} & \textbackslash{}                                                                          \\ \hline
CTAP + $decoder_{phoneme}$                      & 99.0                  & \multicolumn{1}{c|}{\textbf{4.4}}         & \textbf{3.95± 0.066}           & \textbf{3.94 ± 0.008} & \textbf{3.98 ± 0.098} & \multicolumn{1}{c|}{4.9}                  & 3.93 ± 0.076          & 3.79 ± 0.040         & 3.90 ± 0.021          & 95.55                                                                                     \\ \hline
CTAP                                        & 98.2                  & \multicolumn{1}{c|}{4.5}                  & 3.91 ± 0.010 & \textbf{3.94 ± 0.099} & 3.94 ± 0.035          & \multicolumn{1}{c|}{\textbf{4.2}}         & \textbf{4.04 ± 0.057} & \textbf{3.89± 0.040} & 3.92 ± 0.095          & 92.05                                                                                     \\ \hline
\end{tabular}
}
\end{table*}

\subsection{Experimental Step}
In our experiments, we use a speech encoder with a structure similar to Whisper's \cite{radford2023robust} encoder, but with dimensionality reduced to $d$ through a linear layer. The speech encoder consists of 2 convolution layers, a GELU activation function, 6 transformer layers, and a linear layer. The phoneme encoder comprises a convolution layer, a RELU activation function, 4 transformer layers, and a linear layer. The outputs of the speech encoder and phoneme encoder are layer-normalized separately. The prompt encoder is a VAE-based model, consisting of 6 convolutional layers and a SE-ResNet block \cite{hu2018squeeze}. The decoder consists of 6 transformer layers, 5 convolution layers, 4 Tanh activation functions, and a linear layer. The hidden dimension of the encoder and decoder is 512. The models are trained using 8 NVIDIA TESLA V100 32GB GPUs with a batch size of 8 per GPU for 800k steps. Adam\cite{Kingma2014AdamAM} is used as the optimizer with an initial learning rate of 2e-4. 

 \subsection{Datasets}
We combine an internal dataset with the AISHELL-3 dataset\cite{shi2020aishell}, which consists of 215705 recordings (281 hours) of 304 native Mandarin Chinese speakers (73 males and 231 females). We use a test set consisting of 15 minutes of labeled data from each speaker for minimally-supervised TTS and VC tasks. All speech waveforms are sampled at 24kHz and convert to 40-band mel-spectrograms with a frame size of 960 and a hop size of 240. From the perspective of information compression, 40-dimensional features are more suitable for our task.

\subsection{Compared Models}

For the minimally-supervised TTS and VC tasks, we compare our proposed model with four other models, including self-supervised learning methods: {\bf Codec}\cite{Defossez2022HighFN}, {\bf Hubert}\cite{hsu2021hubert}, {\bf Wav2Vec2.0}\cite{baevski2020wav2vec}, and supervised learning methods: the ASR model {\bf Whisper}\cite{radford2023robust}'s encoder. In all models, the embedding values from the codebook are used to replace discrete codes, resulting in continuous-valued inputs and outputs. To ensure fairness, we construct TTS, VC, and ASR systems using the same model structure for all extracted intermediate representations. 

\begin{itemize}
\item All TTS systems are two-stage prediction models ($ text \rightarrow \\ semantic \ prediction \ model  \rightarrow \\ intermediate \ representation + prompt \ embedding \rightarrow  \\ acoustic \ prediction \ model \rightarrow \\ mel$). 

\item All VC systems follow the same structure ($ speech \rightarrow \\ semantic \ extractor \rightarrow \\ intermediate \ representation +  prompt \ embedding \rightarrow \\ acoustic \ prediction \ model \rightarrow \\ mel$). 

\item All ASR systems follow the same structure ($ speech \rightarrow \\ semantic \ extractor \rightarrow \\ intermediate \ representation \rightarrow \\ phoneme \ prediction \ model \rightarrow \\ phoneme$).

\end{itemize}




Specifically, the semantic prediction model, prompt encoder, acoustic prediction model, phoneme prediction model, duration model, and vocoder are the same in all compared models, consistent with our previous work\cite{qiang2023minimally}. The {\bf CTAP w/o decoder} structure is trained with only the $\mathcal{L}_{contrastive}$ and does not include a decoder structure. The {\bf CTAP + $\mathcal{L}_{mse(embed)}$} structure is trained with an additional MSE loss for speech embeddings and phoneme embeddings. 

The {\bf CTAP (Ground Truth)} structure uses intermediate representations extracted from ground truth speech for TTS. The {\bf CTAP + $decoder_{phoneme}$}  is trained using an additional phoneme decoder (with two cross-entropy losses predicting phoneme sequences). The phoneme decoder consists of 6 transformer layers and a linear layer.

\subsection{Test Metrics}
For the minimally-supervised TTS and VC tasks, we conduct all subjective tests using 11 native judgers, with each metric consisting of 20 sentences per speaker. 

The test metrics used in the evaluation include prosody measurement, which involves mean square error for pitch ({\bf MSEP}) to assess prosody similarity against ground-truth speech, word error rate ({\bf WER})(200 sentences), which utilizes an ASR model to transcribe the generated speech, and mean opinion score ({\bf MOS}), which verifies speech quality and similarity in expected speaking prosody and timbre between source speech and synthesized speech. For the ASR task, different methods are compared for phoneme accuracy ({\bf ACC})(200 sentences).

\subsection{Results}
The results of the minimally-supervised TTS task are presented in Table \ref{tab:mos}. The {\bf CTAP}, {\bf CTAP + $\mathcal{L}_{mse(embed)}$}, and {\bf CTAP + $decoder_{phoneme}$} achieve good results in terms of MSEP and WER. The additional MSE loss has little impact on the results. The {\bf CTAP} outperforms the {\bf CTAP w/o decoder}, and the inclusion of a decoder structure and reconstruction loss during training helps to extract intermediate representations that are better suited for frame-level generative tasks. The decoder can be used directly for mel-spectrogram generation in both TTS and VC tasks. In terms of prosody similarity MOS, all systems achieve similar results as they use the same diffusion-based duration model. The {\bf CTAP} and {\bf CTAP + $decoder_{phoneme}$} obtain the best results in terms of speaker similarity MOS by solving the problem of speaker information redundancy in traditional intermediate representations through frame-level contrastive learning. In addition, the pre-trained phoneme encoder addresses the problem of difficult intermediate representation prediction from text to intermediate representation (semantic coding) in two-stage TTS models, and it can be used directly as an effective predictor. All systems achieve similar results in terms of speech quality MOS as they use the same diffusion-based vocoder, but {\bf Codec} and {\bf Hubert}'s discrete processing result in quality loss. 

The results of the VC task are also presented in Table \ref{tab:mos}. All systems achieve good results in terms of WER, prosody similarity MOS, and speech quality MOS. Similarly to the TTS task, {\bf CTAP} obtains the best results in terms of speaker similarity MOS. The phoneme decoder structure in {\bf CTAP + $decoder_{phoneme}$} has a negative impact on performance optimization. Compared to other methods, the intermediate representation extracted by {\bf CTAP} emphasizes semantic content while de-emphasizing paralinguistic information. 

Table \ref{tab:mos} shows the results of the ASR task, all of which achieve acceptable results. The {\bf Hubert} achieves the best results. The presence of the phoneme decoder structure in {\bf CTAP + $decoder_{phoneme}$} during training, which is supervised by the ASR task, results in better performance than {\bf CTAP}. However, the presence of the phoneme decoder structure weakens the ability of contrastive learning, so the final version of {\bf CTAP} is chosen to have the structure, as shown in Fig. \ref{fig:proposed_model}. 

Similar to other contrastive learning-based methods, {\bf CTAP} has better efficiency and significantly reduces computational complexity compared to other pre-training schemes, i.e., it only predicts which frame's phonemes match with which frame's speech. The large number of frames included in each step of our calculations (batch size multiplied by sequence length) is advantageous for contrastive learning.

\section{Conclusions and future work}

In this paper, we propose CTAP for learning speech representation. It solves the problem of dimension explosion and information redundancy in existing speech intermediate representations that lead to prediction difficulties. The method also achieves minimally-supervised TTS, VC, and ASR tasks, which can be extended to downstream tasks involving frame-level generation and recognition. In future work, we will attempt to extract context-independent and sequence-independent frame-level speech representations.

\section{Acknowledgments}
This work is supported by the National Natural Sciencel Foundation of China (No. U23B2053, 62101553).

\vfill\pagebreak

\bibliographystyle{IEEEbib}

\small
\bibliography{strings,refs}

\end{document}